\documentclass[reprint,superscriptaddress,aps] {revtex4-1}
\usepackage{graphicx}
\usepackage{hyperref}
\hypersetup{hidelinks}

\begin{document}


\title{A generalized decay law for particle- and  wave-like thermal phonons}

\author{Zhongwei Zhang}
\affiliation{Center for Phononics and Thermal Energy Science,\\
School of Physics Science and Engineering, Tongji University, 200092
Shanghai, PR China}
\affiliation{China-EU Joint Lab for Nanophononics, Tongji University, 200092 Shanghai, PR China}
\affiliation{Institute of Industrial Science, The University of Tokyo, Tokyo 153-8505, Japan}

\author{Yangyu Guo}
\affiliation{Institute of Industrial Science, The University of Tokyo, Tokyo 153-8505, Japan}

\author{Marc Bescond}
\affiliation{Laboratory for Integrated Micro and Mechatronic Systems, CNRS-IIS UMI 2820, University of Tokyo, Tokyo 153-8505, Japan}

\author{Jie Chen}
\email{jie@tongji.edu.cn}
\affiliation{Center for Phononics and Thermal Energy Science,\\
School of Physics Science and Engineering, Tongji University, 200092
Shanghai, PR China}
\affiliation{China-EU Joint Lab for Nanophononics, Tongji University, 200092 Shanghai, PR China}

\author{Masahiro Nomura}
\email{nomura@iis.u-tokyo.ac.jp}
\affiliation{Institute of Industrial Science, The University of Tokyo, Tokyo 153-8505, Japan}

\author{Sebastian Volz}
\email{volz@iis.u-tokyo.ac.jp}
\affiliation{China-EU Joint Lab for Nanophononics, Tongji University, 200092 Shanghai, PR China}
\affiliation{Laboratory for Integrated Micro and Mechatronic Systems, CNRS-IIS UMI 2820, University of Tokyo, Tokyo 153-8505, Japan}

\date{\today}

\begin{abstract}

Our direct atomic simulations reveal that a thermally activated phonon mode involves a large population of elastic wavepackets. These excitations are characterized by a wide distribution of lifetimes and coherence times expressing particle- and wave-like natures. In agreement with direct simulations, our theoretical derivation yields a generalized law for the decay of the phonon number taking into account coherent effects. Before the conventional exponential decay due to phonon-phonon scattering, this law introduces a delay proportional to the square of the coherence time. This additional regime leads to a moderate increase in the relaxation times and thermal conductivity. This work opens new horizons in the understanding of the origin and the treatment of thermal phonons. 

\end{abstract}

\pacs{Valid PACS appear here}
\maketitle


The legacy of transport physics establishes that thermal phonons of a given mode can be understood as quasi-particles having same lifetime and coherence time. Their behaviors are modeled by Boltzmann transport theory and the phonon-gas model \cite{ZH116,ziman2001}. On the other hand, phonons are defined in essence as vibrational waves. Experimental investigations \cite{ZH606,ZH608,ZH975,RN1631,RN289} demonstrated that the wave nature or the coherence of thermal phonons significantly contributes to thermal transport. For example, Maire $et$ $al.$ \cite{ZH975} could tune thermal conduction by using this coherence in silicon phononic crystals.

Previous studies also demonstrated that the particle- and wave-like thermal phonons are coexisting at elevated temperature \cite{ZH608,ZH608,Latour2017,RN943,RN1729}. In epitaxial oxide superlattices, Ravichandran $et$ $al.$ \cite{ZH608} experimentally observed this mechanism and the predominant behavior depends on temperature and period. However, state-of-the-art theories are failing in simultaneously capturing both wave and particle pictures \cite{ZH116,ziman2001,RN285}. Alternative methods still miss a direct representation of thermal phonon excitations \cite{ZH184,ZH1230,RN1395,RN1438} which is the case when determining phonon lifetimes from anharmonic lattice dynamics \cite{RN1527,RN493,RN1112} and in experimental measurements \cite{ZH1641,RN1414}.

In this letter, we track the real phonon dynamics and extract temporal coherence times and lifetimes by using the wavelet transform of the atomic trajectories during an equilibrium molecular dynamic (MD) simulation. We find that a thermally activated single phonon mode involves a large distribution of excitations with a broad range of coherence times and lifetimes. A theory is proposed to establish the relationship between coherence times and lifetimes, which reveals the unexpected impact of long coherence times on phonon relaxation and thermal conductivity. These conclusions open new insights on the reality of thermally activated phonon modes and their intrinsinc wave-like and coherence behaviors.

Fourier transform has been widely used in phonon-related analysis, since it provides the natural function basis of phonons in the form of monochromatic planewaves $e^{-i\left ( \omega t -\mathbf{k}\cdot \mathbf{ r}\right )}$ \cite{ZH72,ZH117}. Here, $\omega $ and $\mathbf{k}$ are referring respectively to the mode eigenfrequency and wave-vector, and the planewave comprises temporal $t$ and spatial $\mathbf{r}$ dependences. While the amplitude of a single monochromatic planewave propagates, its energy defined as the squared modulus of the amplitude can not. To propagate energy, the extended planewave picture should be upgraded to the one of a spatially and temporally localized wavepacket \cite{ZH35}. In this context, Baker $et$ $al.$ \cite{ZH1637} and Shiomi $et$ $al.$ \cite{ZH1367} proposed wavelet transforms to investigate the propagation of phonon energy, in which the functions of the basis take the form of a Gaussian wavepacket. Modifying this wavelet transform approach and focusing on the temporal information, the expression of the normalized phonon wavelet basis is written as:

\begin{eqnarray}
\centering
\psi_{\omega_{\mathbf{k}s} ,t_{0},\Delta_{\mathbf{k}s} } \left ( t \right )=\pi ^{-\frac{1}{4}}\Delta_{\mathbf{k}s} ^{-\frac{1}{2}}e^{\left [ i\omega_{\mathbf{k}s} \left ( t-t_{0} \right ) \right ]}e^{\left [ -\frac{1}{2}\left ( \frac{t-t_{0}}{\Delta_{\mathbf{k}s}} \right) ^{2} \right ]},
\label{eq_1}
\end{eqnarray}

\noindent where $\omega_{\mathbf{k}s}$ is the angular frequency of mode ${\mathbf{k}s} $, and $\Delta_{\mathbf{k}s} $ defines the wavepacket duration. $t$ corresponds to the time variable, and $t_{0}$ to the position of highest amplitude in the wavepacket and also corresponds to the time evolution in the wavelet space. Inside the wavepacket, planewaves are in phase, the $\Delta_{\mathbf{k}s}$ term in Eq. (\ref{eq_1}) is thus a measure of the temporal coherence of thermal phonons. Here, we define the wavepacket full-width at half-maximum (FWHM) as the coherence time $\tau_{\mathbf{k}s}^{c}=2\sqrt{2ln2}\Delta_{\mathbf{k}s}$. This basis leads to the following wavelet transform:

\begin{eqnarray}
\Lambda \left ( \omega_{\mathbf{k}s},t_{0},\tau_{\mathbf{k}s}^{c}\right )=\int \psi_{\omega_{\mathbf{k}s},t_{0},\tau_{\mathbf{k}s}^{c} } \left ( t \right )F\left (  t\right )dt,
\label{eq_2}
\end{eqnarray}

\noindent where $F\left (  t\right )$ denotes the time dependent dynamical quantity, which is chosen as the phonon modal velocity, $\frac{1}{a}\sum_{b,l}\left [ \mathbf{\dot{u}}_{bl}\left ( t \right )\cdot \mathbf{e}^{\ast } _{b}\left ( \mathbf{k},s \right )\times exp\left ( i\mathbf{k}\cdot \mathbf{R}_{0l} \right )\right ]$, where $\mathbf{\dot{u}}_{bl}\left ( t \right )$ denotes the velocity of the $b$th atom in the $l$th unit cell at time $t$, $a$ the number of cells, $\mathbf{e}^{\ast } \left ( \mathbf{k},s \right )$ refers to the complex conjugate of the eigenvector of mode ${\mathbf{k}s} $, and $\mathbf{R}_{0l} $ is the equilibrium position of the $l$th unit cell. As a control calculation, we validated the relevance of wavelet transform to study phonon properties, i.e. eigenfrequency, temporal coherence and creation/annihilation of wavepackets (See Sec. I in \cite{SM}). We highlight that the wavelet transform successfully provided the same results as those predicted by commonly used spectral energy density (SED) analysis \cite{ZH72,RN1747}.

\begin{figure}[tb]
\includegraphics[width=1.0\linewidth]{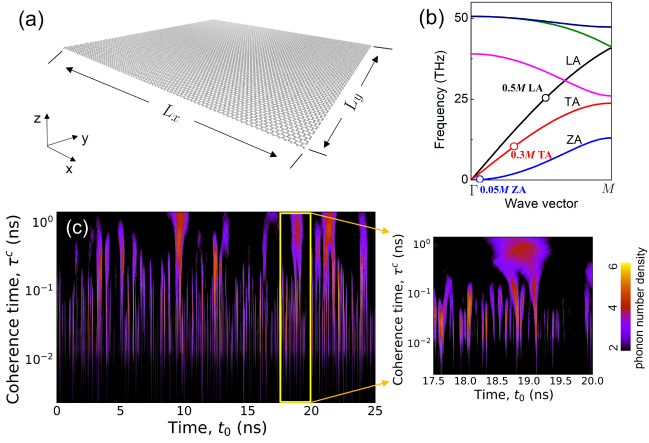}
\caption{Evolution time and coherence time dependent phonon number. (a) Schematic figure of a suspended single-layer graphene. The domain of the graphene system is set to $L_{x}=L_{y}\approx $ 100 nm. The convergence of the results with size was carefully checked. (b) Phonon dispersion of single-layer graphene along $\Gamma(0,0,0) \rightarrow M(0.5,0,0)$. The circles specify the three analyzed modes: $0.05M$ ZA (ZA mode at the kpoint of $0.05\times (0.5,0,0)$), $0.3M$ TA and $0.5M$ LA. (c) Evolution time ($t_{0}$) and coherence time ($\tau _{c}$) dependent phonon number density of the $0.05M$ ZA mode at room temperature. The right-hand side inset highlights the 17.5 - 20.0 ns interval of Figure (c).
}
\label{fig1}
\end{figure}

	 We use classical MD simulations to obtain the real phonon dynamics in the graphene system shown in Fig.\,\ref{fig1}(a) at elevated temperatures. The  C-C interaction is modeled by the optimized Tersoff potential \cite{ZH104}. All MD simulations are performed by implementing the Graphics Processing Units Molecular Dynamics (GPUMD) package \cite{ZH1638} with a timestep of 0.35 fs. To extract a reliable time dependent information, five MD simulations with time duration of 25 ns were carried out. From Eq. (\ref{eq_2}), the time dependent phonon number at a given coherence time, here called phonon number density, $N\left ( \omega_{\mathbf{k}s},t_{0},\tau _{\mathbf{k}s}^{c}  \right )$ can be calculated as $N\left ( \omega_{\mathbf{k}s},t_{0},\tau _{\mathbf{k}s}^{c}  \right )=\frac{1}{2}m\left | \Lambda \left ( \omega_{\mathbf{k}s},t_{0},\tau _{\mathbf{k}s}^{c}  \right ) \right |^{2}/\hbar\omega_{\mathbf{k}s}$, where $m$ refers to the mass of the carbon atom and $\hbar$ is the reduced Planck constant. Wavelet transformation finally provides the mode energy distribution in evolution time and in coherence time. The time dependent phonon number reads $N\left ( \omega_{\mathbf{k}s},t_{0} \right )=\sum_{\tau _{\mathbf{k}s}^{c} }{N}\left ( \omega_{\mathbf{k}s},t_{0},\tau _{\mathbf{k}s}^{c}  \right )$.
	
	Fig.\,\ref{fig1}(c) reports the phonon number density $N\left ( \omega_{\mathbf{k}s},t_{0},\tau _{\mathbf{k}s}^{c}\right )$ derived from Eqs. (\ref{eq_1}) and (\ref{eq_2}) for the $0.05M$ ZA mode. Dark areas indicate the absence of phonon energy and brighter ones represent the apparition of phonon wavepackets. While common understanding stipulates a unique coherence time per mode, we uncover a distribution of coherence times instead. In addition, this distribution is varying with time. The right-hand side inset of Fig.\,\ref{fig1}(c) displays a 2.5 ns time interval of the phonon number density history. It turns out that phonon coherence times of the studied mode cover two orders of magnitude. Sub-populations manifest very long coherence ($\tau _{\mathbf{k}s}^{c}> 300$ ps), i.e. a wave-like feature, whereas other with shorter coherence times ($\tau _{\mathbf{k}s}^{c}$ $<$ 10 ps) can be assimilated to particle-like excitations. 

\begin{figure}[tb]
\includegraphics[width=0.80\linewidth]{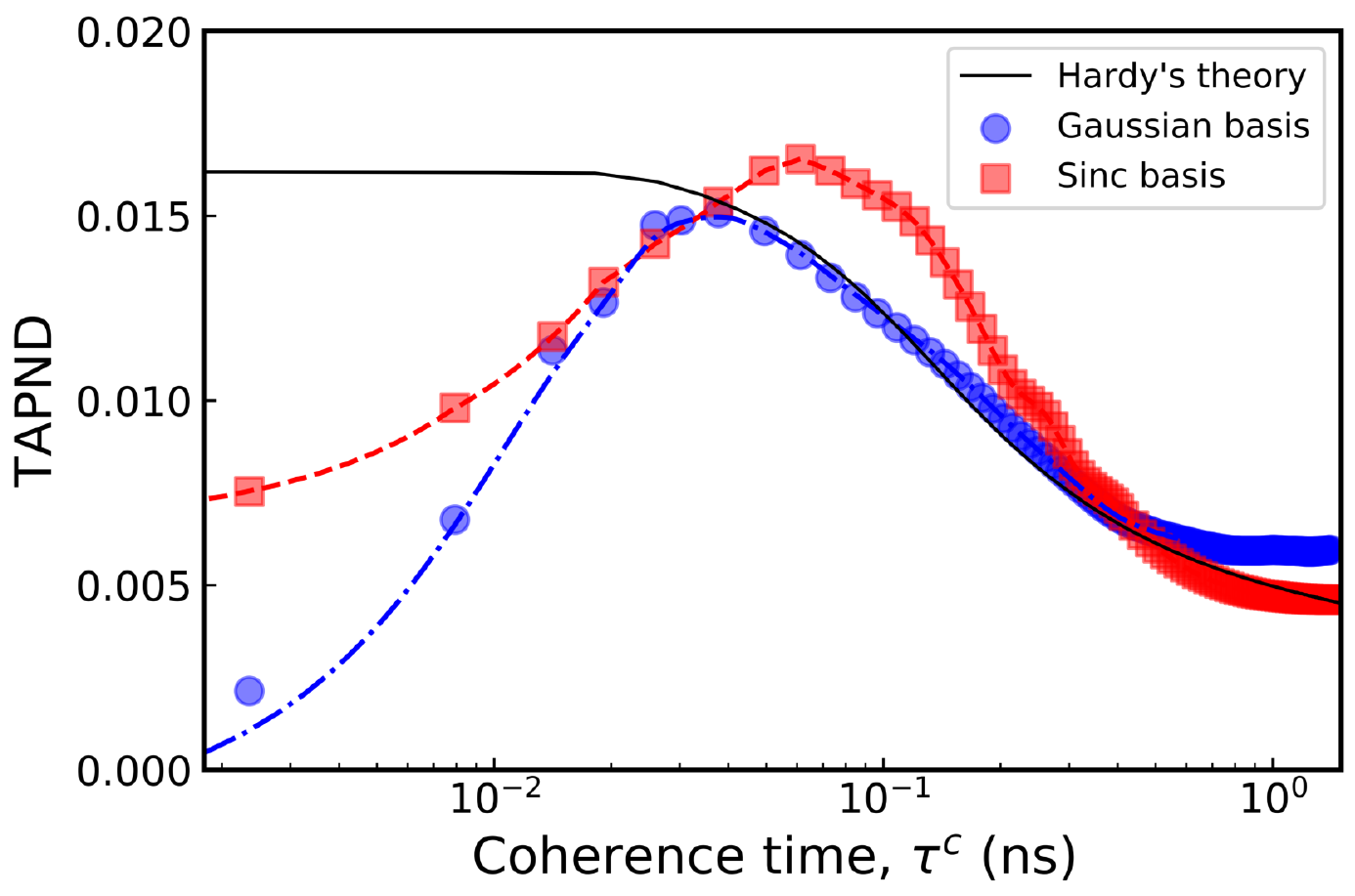}
\caption{Time-averaged phonon number density (TAPND) as a function of coherence time. The density distribution is calculated either with a Gaussian function basis (blue circles) or with a sinus cardinal function one (red squares), for the $0.05M$ ZA mode at room temperature. The solid line is a trend based on Hardy's theory \cite{ZH226}.
}
\label{fig2}
\end{figure}

	The wavepacket distribution can be further investigated by building the time-averaged phonon number density (TAPND) versus coherence time $D\left (\omega _{\mathbf{k}s},\tau _{\mathbf{k}s}^{c} \right )=\frac{1}{N_{t_{0}}}\sum_{t_{0}}\frac{N\left ( \omega_{\mathbf{k}s},t_{0},\tau _{\mathbf{k}s}^{c} \right )}{\sum_{\tau _{\mathbf{k}s}^{c}}N\left ( \omega_{\mathbf{k}s},t_{0},\tau _{\mathbf{k}s}^{c} \right )}$, reported in Fig.\,\ref{fig2}, where $N_{t_{0}}$ denotes the number of terms in the sum. Interestingly, wavepackets follow an unimodal distribution as a function of coherence time, indicating the predominance and the limits in duration of a sub-population centered at $\tau ^{c}\sim$ 40 ps. The TAPND was not observed before, only its averaged on $\tau ^{c}$ obtained by other methods \cite{Latour2017}. Previously, the founding work of Hardy \cite{ZH226} mentioned that the TAPND of a given mode $\mathbf{k}$ should be distributed along an ad hoc Gaussian function, $i.$ $e.$ $D\left (\mathbf{k}, \tau_{\mathbf{k}} ^{c} \right )=e^{ -\frac{1}{4} \left | \mathbf{k}-{\mathbf{K}}\right |^{2} \upsilon _{g}^{2}{\tau_{\mathbf{k}} ^{c}}^{2}}$, where $\mathbf{K}$ denotes the wave vector of a mode interfering with mode $\mathbf{k}$ and $ \upsilon _{g}$ corresponds to the group velocity of mode $\mathbf{k}$. Hardy's prediction fairly agrees with our analysis in the long wavepacket range as highlighted by Fig.\,\ref{fig2}. Disagreement appears in the low coherence time region, presumably because coherence time has to be larger than mode period, which is not taken into account by Hardy's proposition. While a full confirmation for all modes is unreachable, the similar unimodal distribution of the TAPND is also observed for other graphene modes as well as in bulk Silicon (See Sec. V in \cite{SM}).

\begin{figure}[tb]
\includegraphics[width=1.0\linewidth]{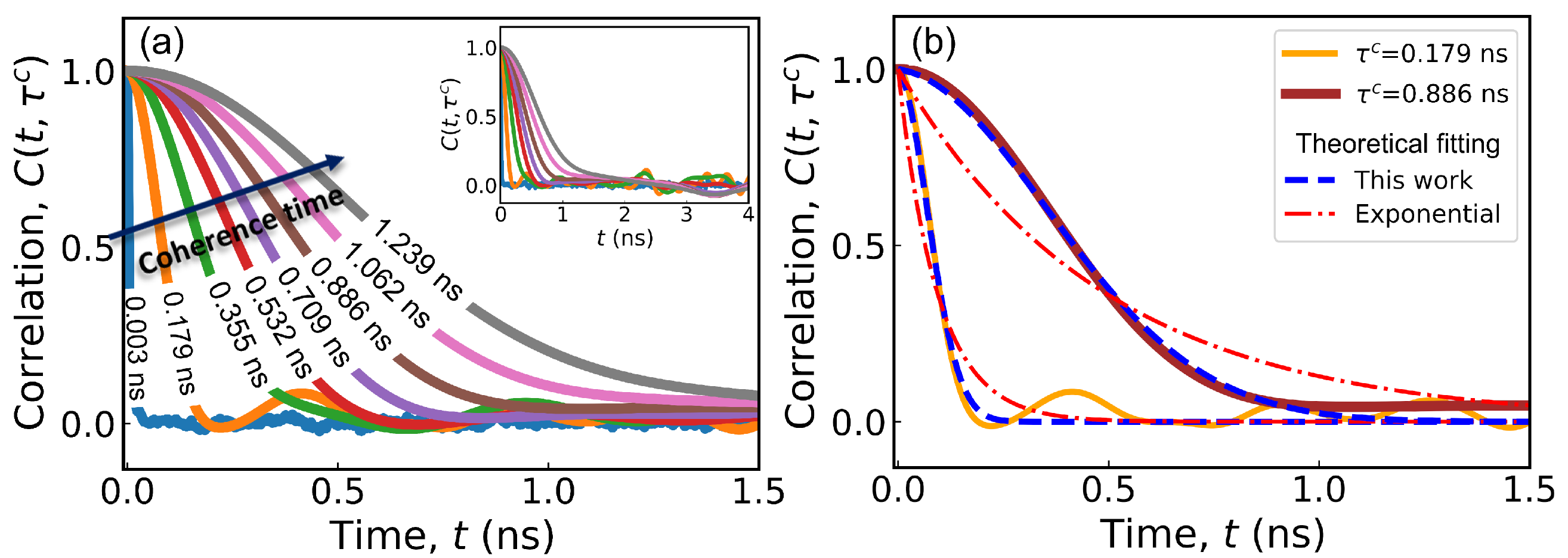}
\caption{Fitting of the phonon number density decay. (a) Phonon number density correlation function $C\left ( t,\tau^{c} \right )$, versus correlation time $t$ for several coherence times $\tau^{c}$. The inset shows the correlation function running over 4ns. (b) Fitting of the MD correlation function (continuous lines) with the conventional exponential decay theory (dashed dot lines) and the theory of this work (dashed lines). 
}
\label{fig3}
\end{figure}

The autocorrelation of the fluctuations of the phonon number density provides a basis for comparison with the conventional description of the phonon mode decay. The autocorrelation function $C\left ( t,\tau _{\mathbf{k}s}^c \right )$ is calculated as $C\left ( t,\tau _{\mathbf{k}s}^c \right )=\left \langle {\Delta N}\left (t,\tau _{\mathbf{k}s}^c   \right ) {\Delta N}\left (0,\tau _{\mathbf{k}s}^c \right ) \right \rangle/\left \langle {\Delta N}\left (0,\tau _{\mathbf{k}s}^c \right ) {\Delta N}\left (0,\tau _{\mathbf{k}s}^c \right ) \right \rangle$, where ${\Delta N}(t_{0},\tau _{\mathbf{k}s}^c)=N\left ( t_{0},\tau _{\mathbf{k}s}^c \right )-\left \langle N(t_{0},\tau _{\mathbf{k}s}^c) \right \rangle_{t_{0}}$. For small coherence time wavepackets, a rapid decay to zero is observed in Fig.\,\ref{fig3}(a) and as coherence time increases, the correlation time does so. In the established knowledge, phonon number decay should be described by an exponential function, $\sim e^{-t/\tau _{\mathbf{k}s} ^{l}}$, where $\tau _{\mathbf{k}s} ^{l}$ is the lifetime for mode $\mathbf{k}s$, when the single-mode relaxation time (SMRT) approximation \cite{ZH1112} applies. Unfolding this decay versus coherence times, the correlations $C\left ( t,\tau _{\mathbf{k}s}^c \right )$ clearly deviate from the exponential trend. This deviation also intensifies with coherence time as illustrated by Fig.\,\ref{fig3}(b). Those results evidence the failure of the SMRT approximation in capturing wavepacket coherence. Note that the effect of this latter differs from the influence of collective mode coupling shown in graphene \cite{Cepellotti2016,RN469} that also invalidates SMRT but preserves the exponential relaxation \cite{Cepellotti2016}. We propose a theoretical insight to clarify why the phonon number density decay should in fact include coherence effects.

	Previously, Hardy \cite{ZH226} demonstrated that the harmonic energy-flux operator $\mathbf{S}$ can be expressed as

\begin{eqnarray}
\mathbf{S}=\frac{1}{V}\sum_{\mathbf{k}s}N_{\mathbf{k}s}\hbar\omega _{\mathbf{k}s}\mathbf{\upsilon }_{\mathbf{k}s},
\label{eq_3}
\end{eqnarray}

\noindent where $V$ is the system volume and $N_{\mathbf{k}s}$ refers to phonon number. If we rewrite Eq. (\ref{eq_3}) in terms of the classical coordinates and consider the fluctuation of phonon energy, the time-dependent phonon number becomes

\begin{eqnarray}
N_{\mathbf{k}s}(t)= \frac{1}{\hbar}\sum_{\mathbf{k}'}p_{\mathbf{k}{s}}\left ( t \right )q_{\mathbf{k}'{s}}^{\ast }\left ( t \right ),
\label{eq_4}
\end{eqnarray}

\noindent where, $p_{\mathbf{k}s}\left ( t \right )$ and $q_{\mathbf{k}s}\left ( t \right )$ denote the time-dependent normal mode momentum and displacement for mode $\mathbf{k}s$. $\ast$ indicates the conjugate form. Here, we ignored the terms $s\neq {s}'$ that are rapidly oscillating and yield negligible time average. Because of phonon-phonon scattering, normal mode coordinate decays with time \cite{RN1527}, i.e. $q_{\mathbf{k}s}\left ( t \right )=q_{\mathbf{k}s}\left (0 \right )e^{-\Gamma _{\mathbf{k}s}t-i\omega_{\mathbf{k}s} t}$, where $\Gamma _{\mathbf{k}s}$ represents the mode linewidth. Accordingly, the time-dependent momentum can be expressed as $p_{\mathbf{k}{s}}\left ( t \right )=p_{\mathbf{k}{s}}\left ( 0 \right )e^{-\Gamma _{\mathbf{k}{s}}t-i\omega_{\mathbf{k}s} t}$. Consequently, the time evolution of the phonon number is obtained as

\begin{eqnarray}
N_{\mathbf{k}s}\left ( t \right )=\sum_{\mathbf{k}'}\xi_{\mathbf{k}\mathbf{k}'s}e^{-\gamma _{\mathbf{k}\mathbf{k}'s}t-i\Delta\omega_{\mathbf{k}\mathbf{k}'s} t},
\label{eq_5}
\end{eqnarray}

\noindent where, $\xi_{\mathbf{k}\mathbf{k}'s}=p_{\mathbf{k}s}\left (0 \right )q_{\mathbf{k}'{s}}^{\ast }\left ( 0 \right )/\hbar$, $\gamma _{\mathbf{k}\mathbf{k}'s}=\Gamma _{\mathbf{k}s}+\Gamma _{\mathbf{k}'{s}}$ and $\Delta\omega_{\mathbf{k}\mathbf{k}'s}= \omega_{\mathbf{k}s}-\omega_{\mathbf{k}'{s}}$. It should be noted that the summation over $\mathbf{k}',\mathbf{k}\neq {\mathbf{k}}'$, can be understood as the interference of planewaves defined by $\mathbf{k}{s}$ and $\mathbf{k}'{s}$, which are forming wavepackets. As indicated by Fig.\,\ref{fig1}(c), we can assume that a single wavepacket appears at a given time. Considering that this wavepacket is resulting from its specific and restricted frequency interval $\left [ \omega_{\mathbf{k}s}-\frac{\Omega_{\mathbf{k}s} }{2},\omega_{\mathbf{k}s}+\frac{\Omega_{\mathbf{k}s} }{2} \right ]$ where density of states and linewidth are nearly constant, a further step can be taken

\begin{small}
\begin{eqnarray}
N_{\mathbf{k}s}\left ( t \right )\approx  \bar{\xi}_{\mathbf{k}s}e^{-\bar{\gamma} _{\mathbf{k}s}t}\bar{g}(\omega_{\mathbf{k}s})\int_{-\frac{\Omega_{\mathbf{k}s} }{2}}^{+\frac{\Omega_{\mathbf{k}s} }{2}}e^{-i\Delta\omega_{\mathbf{k}\mathbf{k}'s} t}d\Delta\omega_{\mathbf{k}\mathbf{k}'s},
\label{eq_6}
\end{eqnarray}
\end{small}

\noindent where $\bar{\xi}_{\mathbf{k}s}$ and $\bar{\gamma} _{\mathbf{k}s}$ correspond to the averaged properties over the wavepacket frequency interval and $\bar{g}(\omega_{\mathbf{k}s})$ refers to the density of states. Estimation of the integral in Eq. (\ref{eq_6}) yields (see Sec. II in \cite{SM} for derivation)

\begin{eqnarray}
N_{\mathbf{k}s}\left ( t \right )= 2\bar{\xi}_{\mathbf{k}s}\bar{g}(\omega_{\mathbf{k}s})e^{-\bar{\gamma} _{\mathbf{k}s}t}
 \frac{\sin \pi \Omega_{\mathbf{k}s} t}{t}.
 \label{eq_7}
\end{eqnarray}

\noindent This derivation indicates that at time $t$ a phonon wavepacket with properties $\bar{\gamma} _{\mathbf{k}s}$ and $\Omega_{\mathbf{k}s}$ is contributing to the phonon number $N_{\mathbf{k}s}\left ( t \right )$. Moreover, this phonon wavepacket should take the form of a sinus cardinal function, which is close to the Gaussian function but with a longer tail (See Fig. S2 in \cite{SM}). These two functions indeed have the same form in the vicinity of the origin, and the trends of TAPND coincide fairly well with each other as reported in Fig.\,\ref{fig2}. To analytically investigate the wave effect on phonon decay, the autocorrelation function of the phonon number density can be further derived as (see Sec. III in \cite{SM} for complete derivation)

\begin{eqnarray}
C\left ( t,\tau _{\mathbf{k}s}^{c} \right )= e^{-\frac{t}{2\tau _{\mathbf{k}s}^{l} }}e^{-4ln2\frac{t^{2}}{{\tau _{\mathbf{k}s}^{c}}^{2}}}, 
 \label{eq_8}
\end{eqnarray}

\noindent where, the two terms on right-hand side correspond to the two distinct behaviors of thermal phonons, i.e. particle-like and wave-like. The particle-like part remains an exponentially decaying function, and the wave-like one appears as a quadratic gaussian term. The autocorrelation of Eq. (\ref{eq_8}) can be factorized into:

\begin{eqnarray}
C\left ( t,\tau _{\mathbf{k}s}^{c} \right )=e^{-4ln2\frac{\left ( t +\tau_{w}\right )^{2}}{{\tau _{\mathbf{k}s}^{c}}^{2}}} e^{\frac{\tau_{w}}{4\tau _{\mathbf{k}s}^{l} }},
 \label{eq_9}
\end{eqnarray}

\noindent where $\tau_{w}={\tau _{\mathbf{k}s}^{c}}^{2}/\left ( \tau _{\mathbf{k}s}^{l}16ln2 \right )$ evaluates the effect of coherence on phonon decay. This latter equation reveals the general form of a gaussian decay for the phonon number in which the very short time regime $t\ll \tau_{w}$ yields strong correlation $C\left ( t,\tau _{\mathbf{k}s}^{c} \right ) \approx 1$ as seen in Fig.\,\ref{fig3}(a) (line with $\tau _{\mathbf{k}s}^{c}=1.239$ ns, $t< 0.5$ ns), and intermediate times $t<\tau _{w}$ lead back to the usual phonon-phonon scattering decay $C\left ( t,\tau _{\mathbf{k}s}^{c} \right )= e^{-t/2\tau _{\mathbf{k}s}^{l}}$ after Taylor expansion. The longer time interval $t>\tau _{w}$ brings a gaussian decay $C\left ( t,\tau _{\mathbf{k}s}^{c} \right )=e^{-4ln2t^{2}/{\tau _{\mathbf{k}s}^{c}}^{2}}e^{\tau_{w}/{4\tau _{\mathbf{k}s}^{l}}}$. The first short time regime is due to the build-up of the wavepacket and is delaying the relaxation compared to the conventional exponential decay. From Eq. (\ref{eq_8}), we can simultaneously obtain $\tau _{\mathbf{k}s}^{l} $ and $\tau _{\mathbf{k}s}^{c} $ for a given mode by fitting the autocorrelations of Fig.\,\ref{fig3}. As shown in Fig.\,\ref{fig3}(b), the disagreement between MD results and the exponential fitting is larger compared to the one resulting from the predictions of Eq. (\ref{eq_8}).

\begin{figure}[t]
\includegraphics[width=1.0\linewidth]{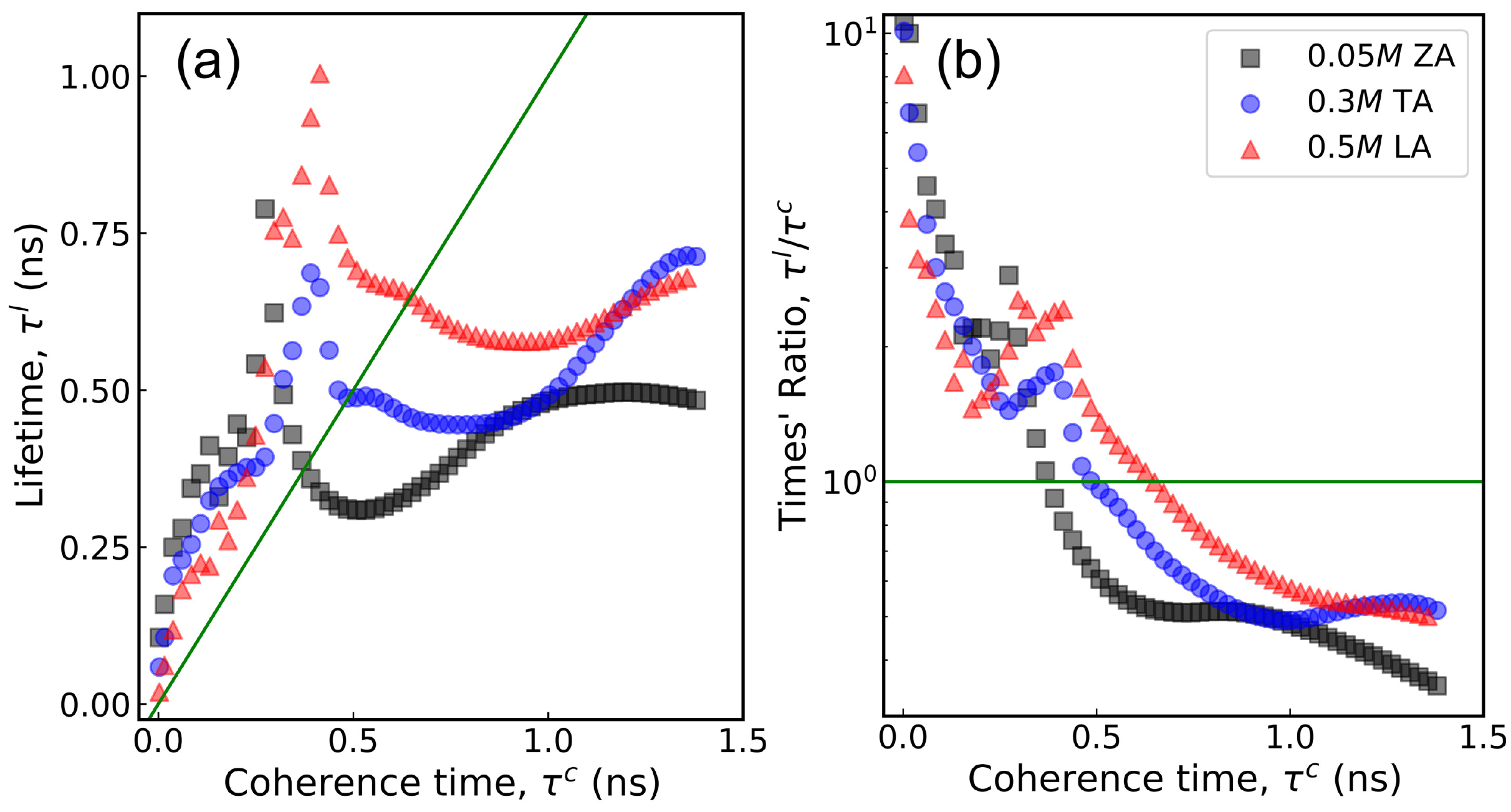}
\caption{Lifetime and coherence time in graphene. (a) Fitted lifetime and coherence time in graphene, and (b) ratio between lifetime and coherence time for modes $0.05M$ ZA, $0.3M$ TA and $0.5M$ LA at room temperature. The green line indicates the lifetimes = coherence times condition. Above the line, phonons show particle-like nature properties with lifetimes $>$ coherence times. In contrast, below the line, phonons have wave-like nature properties with lifetimes $<$ coherence times.
}
\label{fig4}
\end{figure}

Fig.\,\ref{fig4} shows the fitted lifetimes and coherence times according to the proposed theory. Obviously, for short wavepackets, i.e. small coherence times, phonons exhibit prominent particle-like behaviors but with long lifetimes. As coherence time increases, the time spread of the phonon wavepackets becomes longer, with pronounced wave behaviors as illustrated in Fig.\,\ref{fig4}(b). The lifetime follows a non-monotonic dependence on coherence time depicted in Fig.\,\ref{fig4}(a) with a transition point around the coherence time of 0.45 ns. Larger phonon wavepackets ($\tau _{\mathbf{k}s}^{c}>$ 0.45 ns) possess shorter lifetimes, while a set of wavepackets at $\sim$ 0.45 ns survive on longer periods. Note that phonons are almost monotonically transiting from the particle-like nature to the wave-like nature with increasing coherence time as shown by the time ratio in Fig.\,\ref{fig4}(b). In addition, this transition is also frequency dependent as low frequency phonons in the $0.05M$ ZA mode display wave-like behaviors at shorter coherence times.

\begin{figure}[b]
\includegraphics[width=1.0\linewidth]{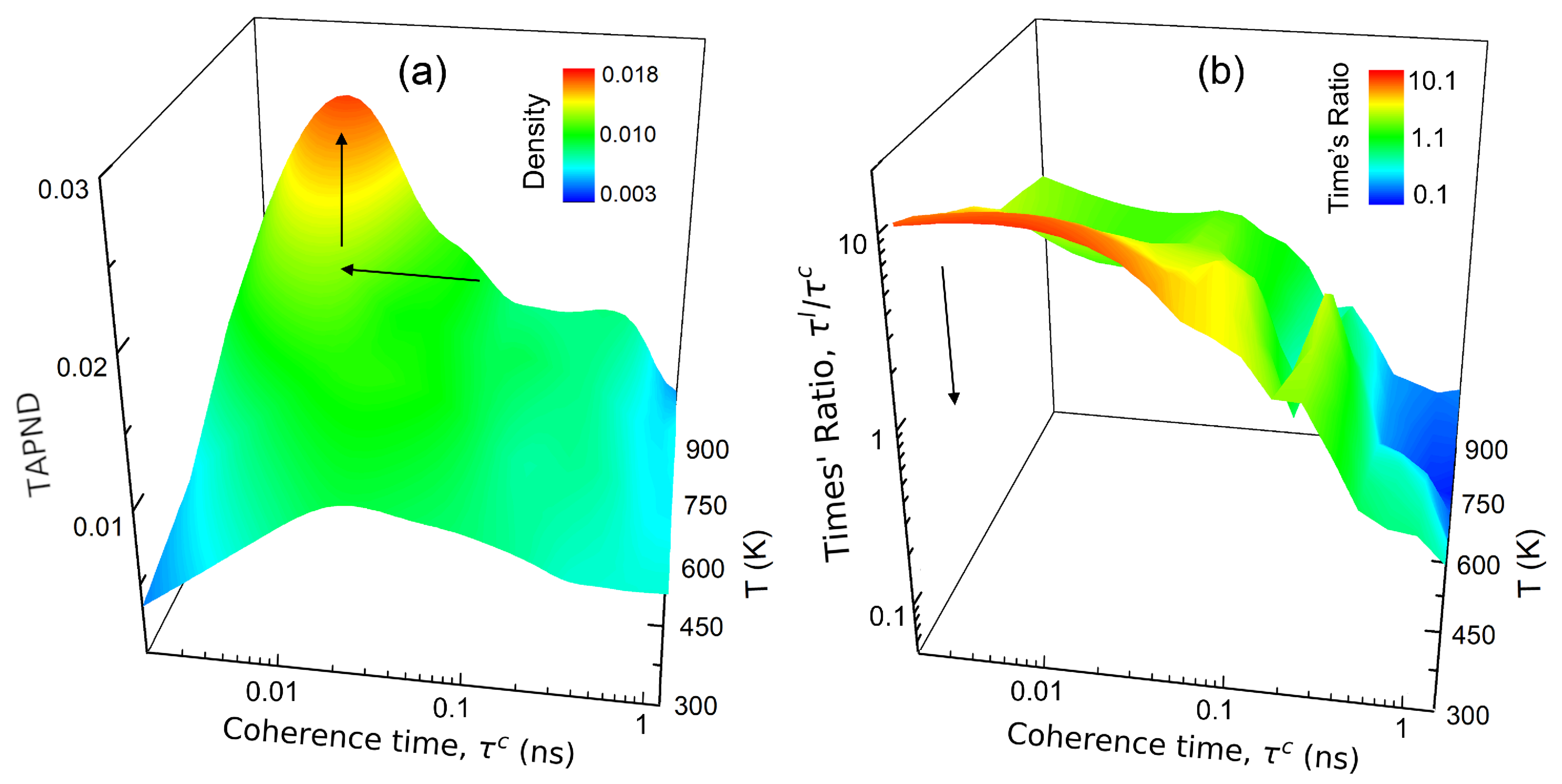}
\caption{Temperature effect on the phonon number density. (a) Time-averaged phonon number density (TAPND) $D\left (\omega,\tau _{c}   \right )$ versus coherence time $\tau ^{c} $ as a function of temperature for the $0.05M$ ZA mode. The arrows indicate the shift of the peak towards lower $\tau ^{c} $ values and a higher density with increasing temperature. (b) Times' ratio $\tau ^{l}/\tau ^{c} $ versus coherence time $\tau ^{c} $ as a function of temperature for the $0.05M$ ZA mode. The arrow indicates the shift of the ratio $\tau ^{l}/\tau ^{c} $ towards lower densities with temperature.
}
\label{fig5}
\end{figure}

The influence of temperature on the wavepacket coherence time and lifetime was also studied in Fig.\,\ref{fig5}. As temperature increases, phonon-phonon scattering intensifies and the particle nature of phonons, i.e. the contribution of short coherence time wavepackets, becomes more obvious. As reported in Fig.\,\ref{fig5}(a), this trend is evidenced by a shift of the density peak to lower coherence times and by the gradual suppression of long wavepackets. Moreover, the decrease of phonon lifetimes with temperature can also be observed in the reduction of the times' ratio $\tau ^{l}/\tau ^{c} $ with temperature.

 The mean autocorrelation is calculated as $\sum_{\tau _{\mathbf{k}s}^{c}}D\left (\tau _{\mathbf{k}s}^{c} \right )C\left ( t,\tau _{\mathbf{k}s}^{c} \right )$, in which $D\left (\tau _{\mathbf{k}s}^{c} \right )$ is the TAPND for mode $\mathbf{k}s$ and the autocorrelation $C\left ( t,\tau _{\mathbf{k}s}^{c} \right )$ is corrected by including coherent effects. By fitting the mean autocorrelation with an exponentially decaying function, the corrected lifetimes and thermal conductivities were computed and compared to the non-corrected ones (See Fig. S4). The corrected lifetimes are larger than the ones obtained from the usual description especially for low frequency phonons. And the coherence correction at room temperature in thermal conductivity introduces a deviation of 16.7 $\%$, indicating that wavepacket time spread plays a significant role on phonon transport but was ignored before. 

The wavelet transform calculations and the proposed phonon decay theory (Eq. (\ref{eq_8})) prove that a single thermally activated phonon modes includes excitations with a broad range of lifetimes and coherence times. The commonly used SMRT theory is based on a single phonon lifetime per mode and the assumption of exponential decay to fit the phonon number autocorrelation function. This letter proposes a step forward, by unfolding the phonon number over the coherence times and rewriting the autocorrelation function of this number as a combination of coherence time dependent correlation functions.

Beyond the here investigated phonon propagation, the wave-like behavior of phonons should also be critical in the treatment of surface phonon scattering and propagation in nanostructures, superlattices and interfaces. Uncovering here the unexpected content of thermal phonon excitations leads to new perspectives in the understanding of thermal coherence and the essence of a black body.

\textit{Acknowledgments}---
This project is supported in part by the grants from the National Natural Science Foundation of China (Grant No. 11890703), and Science and Technology Commission of Shanghai Municipality (Grant Nos. 19ZR1478600, 18JC1410900 and 17ZR1448000). J. C. acknowledges support from the National Youth 1000 Talents Program in China. This work is partially supported by CREST JST (No. JPMJCR19Q3) and Kakenhi (Nos. 15H05869 and 17H02729). Z. Z. gratelfully acknowledge financial support from China Scholarship Council.

\bibliographystyle{apsrev4-2}
\bibliography{library}

\end{document}



\title{Supplemental Material for ``A generalized decay law for particle- and  wave-like thermal phonons"}

\author{Zhongwei Zhang}
\affiliation{Center for Phononics and Thermal Energy Science,\\
School of Physics Science and Engineering, Tongji University, 200092
Shanghai, PR China}
\affiliation{China-EU Joint Lab for Nanophononics, Tongji University, 200092 Shanghai, PR China}
\affiliation{Institute of Industrial Science, The University of Tokyo, Tokyo 153-8505, Japan}

\author{Yangyu Guo}
\affiliation{Institute of Industrial Science, The University of Tokyo, Tokyo 153-8505, Japan}

\author{Marc Bescond}
\affiliation{Laboratory for Integrated Micro and Mechatronic Systems, CNRS-IIS UMI 2820, University of Tokyo, Tokyo 153-8505, Japan}

\author{Jie Chen}
\email{jie@tongji.edu.cn}
\affiliation{Center for Phononics and Thermal Energy Science,\\
School of Physics Science and Engineering, Tongji University, 200092
Shanghai, PR China}
\affiliation{China-EU Joint Lab for Nanophononics, Tongji University, 200092 Shanghai, PR China}

\author{Masahiro Nomura}
\email{nomura@iis.u-tokyo.ac.jp}
\affiliation{Institute of Industrial Science, The University of Tokyo, Tokyo 153-8505, Japan}

\author{Sebastian Volz}
\email{volz@iis.u-tokyo.ac.jp}
\affiliation{China-EU Joint Lab for Nanophononics, Tongji University, 200092 Shanghai, PR China}
\affiliation{Laboratory for Integrated Micro and Mechatronic Systems, CNRS-IIS UMI 2820, University of Tokyo, Tokyo 153-8505, Japan}

\date{\today}

\maketitle

\section{\label{sec:level1}Validation of wavelet transform}

\begin{figure}[h]
\renewcommand\thefigure{S\arabic{figure}}
\includegraphics[width=0.80\linewidth]{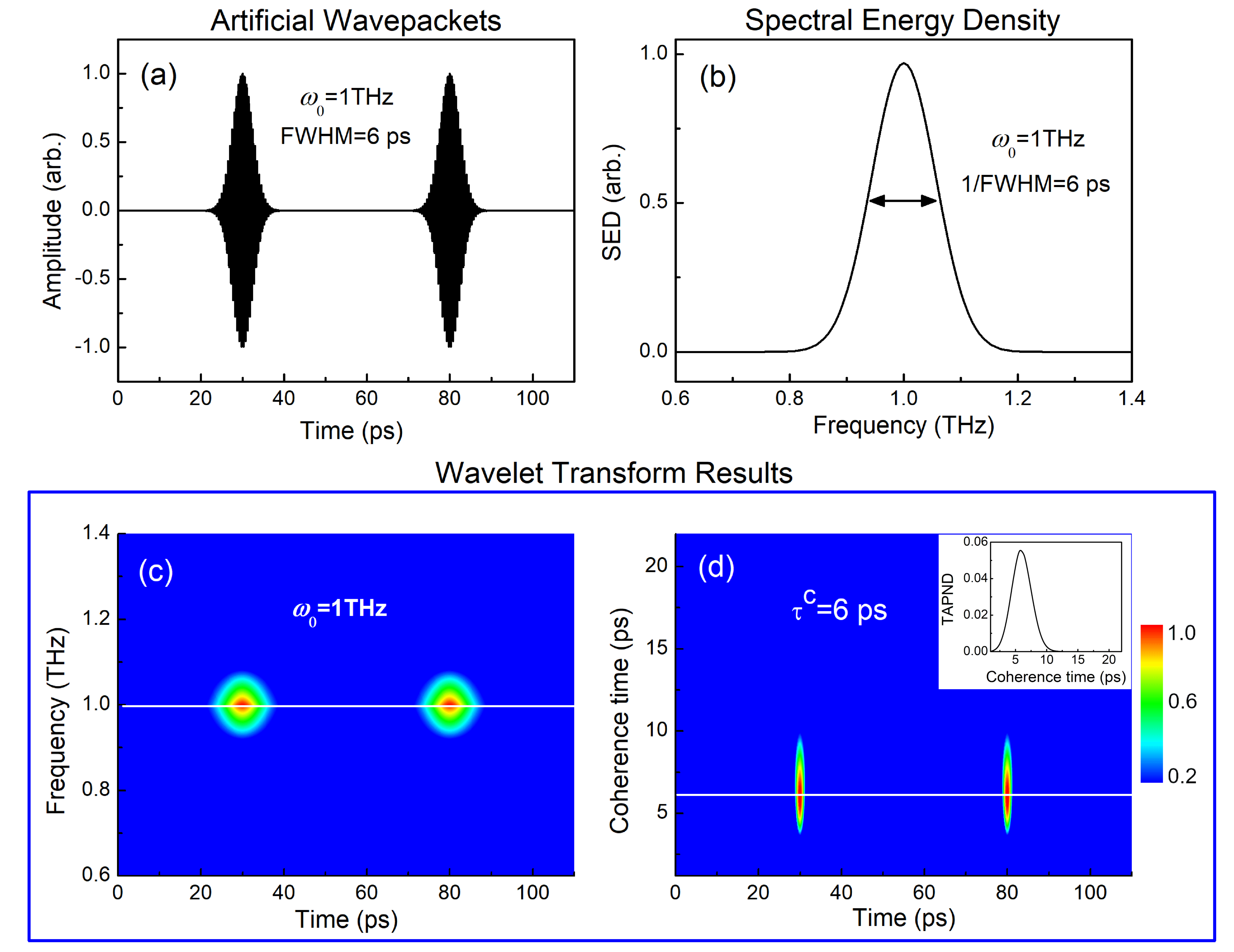}
\caption{Validation of the wavelet transform in the study of phonon properties. (a) Two artificial wavepackets generated based on Eq. 1 with frequency 1 THz and full-width at half-maximum (FWHM) 6 ps, triggered at times 30 ps and 80 ps. (b) Spectral energy density (SED) of the artificial wavepackets in (a). (c) Frequency versus evolution time obtained from wavelet transform. (d) Coherence time versus evolution time obtained from wavelet transform. The inset of (d) is the time averaged phonon number density (TAPND) versus coherence time. The color bar indicates the normalized phonon number density.
}
\label{bench}
\end{figure}

A benchmark is proposed to illustrate the reliability of the wavelet transform approach in the study of phonon properties. Firstly, artificial wavepackets are generated with forms based on Eq. 1 of the main text. As shown in Fig.\,\ref{bench}(a), two artificial wavepackets are triggered at times 30 ps and 80 ps. Other parameters are set as 1 THz eigenfrequency and 6 ps full-width at half-maximum (FWHM). Usually, the spectral energy density method is used to study phonon properties, especially phonon lifetime and eigenfrequency [1]. The calculated times in SED, which is always expressed as the reverse of the SED FWHM, is also the temporal spreading of the wavepacket and thus the coherence times. By using the SED method, we precisely retrieve the same set of parameters than the ones of the artificial wavepackets.

By varying frequency and performing the wavelet transform in Eqs. (1) and (2), Fig.\,\ref{bench}(c) reports the frequency versus evolution time for the artificial wavepackets. Obviously, we can accurately obtain the frequency, the creation and annihilation times of wavepackets from the input information. On the other hand, by varying the coherence time in Eq. 1 and performing the wavelet transform, we can extract the information of coherence time versus evolution time. As shown in Fig.\,\ref{bench}(c), the obtained mean coherence time agrees well with the initial set of parameters. Moreover, the inset of Fig.\,\ref{bench}(d) shows the TAPND as a function of coherence time. Firstly, the peak position at $\sim$ 6 ps agrees well with the set of coherence time in the artificial wavepackets. In addition, the broadening of this peak of 3.9 ps is several order of magnitudes smaller than the studied coherence time of phonon wavepackets in graphene, indicating a sufficient resolution in coherence time of the wavelet transform. Therefore it can be concluded that the wavelet transform can precisely obtain the phonon properties, i.e. eigenfrequency, coherence time and also wavepackets creation and annihilation dynamics, which agrees well with the commonly used SED method.

\section{\label{sec:level1}Heat flux and Time dependent phonon number}

In the following, we focus on the harmonic energy flux. Thus, the simplified energy-flux operator for mode ${\mathbf{k}s}$ is defined as in Reference [2]

\begin{eqnarray}
\mathbf{S}=-\frac{1}{2V}\sum_{\mathbf{k}{\mathbf{k}}',s{s}'}\left ( a_{\mathbf{k}s}+a_{\mathbf{-k}s}^{\dagger } \right )\times \left ( a_{{\mathbf{k}}'{s}'}-a_{{\mathbf{-k}}'{s}'}^{\dagger } \right )\hbar\omega _{\mathbf{k}s}\mathbf{\upsilon }_{\mathbf{k}s},
\label{t_8}
\end{eqnarray}

\noindent where, $a_{\mathbf{k}s}$ and $a_{\mathbf{k}s}^{\dagger }$ respectively denote the phonon annihilation and creation operators. $V$ is the volume of system, and $\omega _{\mathbf{k}s}$ and $\mathbf{\upsilon }_{\mathbf{k}s}$ are the eigenfrequency and group velocity for mode ${\mathbf{k}s}$.

The terms in Eq. (\ref{t_8}) with $s\neq {s}'$ contain contributions from modes with significantly different frequencies. Thus, their interferences are rapidly oscillating and become negligible after time averaging, accordingly these terms will be neglected. For the terms $s= {s}'$ and $\mathbf{k}\neq{\mathbf{k}}'$, the frequencies can be close and the resulting interferences can be kept in the further steps. The corresponding classical version of Eq. (\ref{t_8}) reads

\begin{eqnarray}
\mathbf{S}=\frac{1}{V}\sum_{\mathbf{k}\mathbf{k}',s}p_{\mathbf{k}s}q_{\mathbf{k}'{s}}^{\ast } \omega _{\mathbf{k}s}\mathbf{\upsilon }_{\mathbf{k}s},
\label{t_9}
\end{eqnarray}

\noindent where $p_{\mathbf{k}s}$ and $q_{\mathbf{k}{s}}$ are the modal momentum and velocity for mode $\mathbf{k}{s}$. They can be obtained as 

\begin{eqnarray}
\begin{matrix}
\mathbf{P}(\mathbf{x}_{i})=N^{-1/2}\sum_{\mathbf{k}s}p_{\mathbf{k}s}\mathbf{e}_{\mathbf{k}s}e^{i{\mathbf{k}\cdot {\mathbf{x}_{i}}}} \\ 
\mathbf{Q}(\mathbf{x}_{i})=N^{-1/2}\sum_{\mathbf{k}s}q_{\mathbf{k}s}^{\ast }\mathbf{e}_{\mathbf{k}s}e^{i{\mathbf{k}\cdot {\mathbf{x}_{i}}}} 
\end{matrix}
\label{t_7}
\end{eqnarray}

\noindent here, $\mathbf{P}(\mathbf{x}_{i})$ and $\mathbf{Q}(\mathbf{x}_{i})$ are respectively momentum and position operators of the $ith$-particle. $\mathbf{e}_{\mathbf{k}s}$ is the eigenvector of mode ${\mathbf{k}s}$, and $N$ is the total number of particles. 

When introducing the time dependence into the energy flux, this latter can be expressed as as a function of the products of the modal momentum ($p_{\mathbf{k}s}\left ( t \right )$) and displacement ($q_{\mathbf{k}'{s}}\left ( t \right )$) fluctuations,

\begin{eqnarray}	
	\mathbf{S}\left ( t \right )=\frac{1}{V}\sum_{\mathbf{k}\mathbf{k}',s}p_{\mathbf{k}s}\left ( t \right )q_{\mathbf{k}'{s}}^{\ast}\left ( t \right ) \omega _{\mathbf{k}s}\mathbf{\upsilon }_{\mathbf{k}s}.
\label{t_10}
\end{eqnarray}

Considering that the phonon population can be quantified by $N_{\mathbf{k}s}$, the harmonic heat flux reduces to

\begin{eqnarray}
\mathbf{S}=\frac{1}{V}\sum_{\mathbf{k}s}N_{\mathbf{k}s}\hbar\omega _{\mathbf{k}s}\mathbf{\upsilon }_{\mathbf{k}s}.
\label{t_11}
\end{eqnarray}

The time dependent phonon number quantity $N_{\mathbf{k}s}(t)$ can be obtained as

\begin{eqnarray}
N_{\mathbf{k}s}(t)=\frac{1}{\hbar} \sum_{\mathbf{k}'}p_{\mathbf{k}s}\left ( t \right )q_{\mathbf{k}'{s}}^{\ast }\left ( t \right ).
\label{t_12}
\end{eqnarray}

\noindent On this basis, we can study the fluctuations of the phonon number $N_{\mathbf{k}s}(t)$ from the dynamics of atomic motions.

Moreover, including phonon-phonon scattering in the decay of normal mode coordinates yields [3], 

\begin{eqnarray}
q_{\mathbf{k}'{s}}\left ( t \right )=q_{\mathbf{k}'{s}}\left ( 0 \right )e^{-\Gamma _{\mathbf{k}'{s}}t-i\omega_{\mathbf{k}'s} t},
\label{t_13}
\end{eqnarray}

\noindent where $\Gamma _{\mathbf{k}'s}$ is the linewidth of the mode ${\mathbf{k}'s}$ corresponding to phonon-phonon scattering. The term $e^{-i\omega_{\mathbf{k}'s} t}$ indicates the monochromaticity of the eigenmode, while $e^{-\Gamma _{\mathbf{k}'s}t}$ term corresponds to the decay of the vibration amplitude. The conjugate expression is defined as $q_{\mathbf{k}'{s}}^{\ast }\left ( t \right )=q_{\mathbf{k}'{s}}^{\ast }\left ( 0 \right )e^{-\Gamma _{\mathbf{k}'{s}}t+i\omega_{\mathbf{k}'s} t}$. This type of decay has been verified in inelastic neutron scattering measurements.

Accordingly, we assume that the time dependent modal momentum is 

\begin{eqnarray}
p_{\mathbf{k}s}\left ( t \right )=p_{\mathbf{k}s}\left (0 \right )e^{-\Gamma _{\mathbf{k}s}t-i\omega_{\mathbf{k}s} t}.
\label{t_14}
\end{eqnarray}

\noindent Consequently, the dynamic of the phonon number is obtained as

\begin{eqnarray}
N_{\mathbf{k}s}\left ( t \right )= \sum_{\mathbf{k}'}\xi_{\mathbf{k}\mathbf{k}'s}e^{-\gamma _{\mathbf{k}\mathbf{k}'s}t-i\Delta\omega_{\mathbf{k}\mathbf{k}'s} t},
\label{t_15}
\end{eqnarray}

\noindent where the variables in the above equation are $\xi_{\mathbf{k}\mathbf{k}'s}=p_{\mathbf{k}s}\left (0 \right )q_{\mathbf{k}'{s}}^{\ast }\left ( 0 \right )/\hbar$, $\gamma _{\mathbf{k}\mathbf{k}'s}=\Gamma _{\mathbf{k}s}+\Gamma _{\mathbf{k}'{s}}$ and $\Delta\omega_{\mathbf{k}\mathbf{k}'s}= \omega_{\mathbf{k}s}-\omega_{\mathbf{k}'{s}}$. 

It should also be noted that the summation over $\mathbf{k}'{s}$ can be understood as the interference of $\mathbf{k}'{s}$ and $\mathbf{k}{s}$ planewaves. During the phonon dynamics, the interference between different planewaves result in the generation of a phonon wavepacket. In addition, we can assume that the frequency $\omega_{\mathbf{k}'s}$ of the interfering planewaves is in the range of $\left [ \omega_{\mathbf{k}s}-\frac{\Omega_{\mathbf{k}s} }{2},\omega_{\mathbf{k}s}+\frac{\Omega_{\mathbf{k}s} }{2} \right ]$. Moreover, to shift from a discrete summation to a continuous one, we also introduce the density of states for mode $\omega_{\mathbf{k}'s}$, i.e. $g(\omega_{\mathbf{k}'s})$. Accordingly, Eq. (\ref{t_15}) can be rewritten in the integral form as

\begin{eqnarray}
N_{\mathbf{k}s}\left ( t \right )\approx  \int_{\omega_{\mathbf{k}s}-\frac{\Omega_{\mathbf{k}s} }{2}}^{\omega_{\mathbf{k}s}+\frac{\Omega_{\mathbf{k}s} }{2}}\xi_{\mathbf{k}\mathbf{k}'s}e^{-\gamma _{\mathbf{k}\mathbf{k}'s}t-i\Delta\omega_{\mathbf{k}\mathbf{k}'s} t}g(\omega_{\mathbf{k}'s})d\omega_{\mathbf{k}'s}.
\label{t_16}
\end{eqnarray}

\begin{figure} 
\renewcommand\thefigure{S\arabic{figure}}
\includegraphics[width=0.80\linewidth]{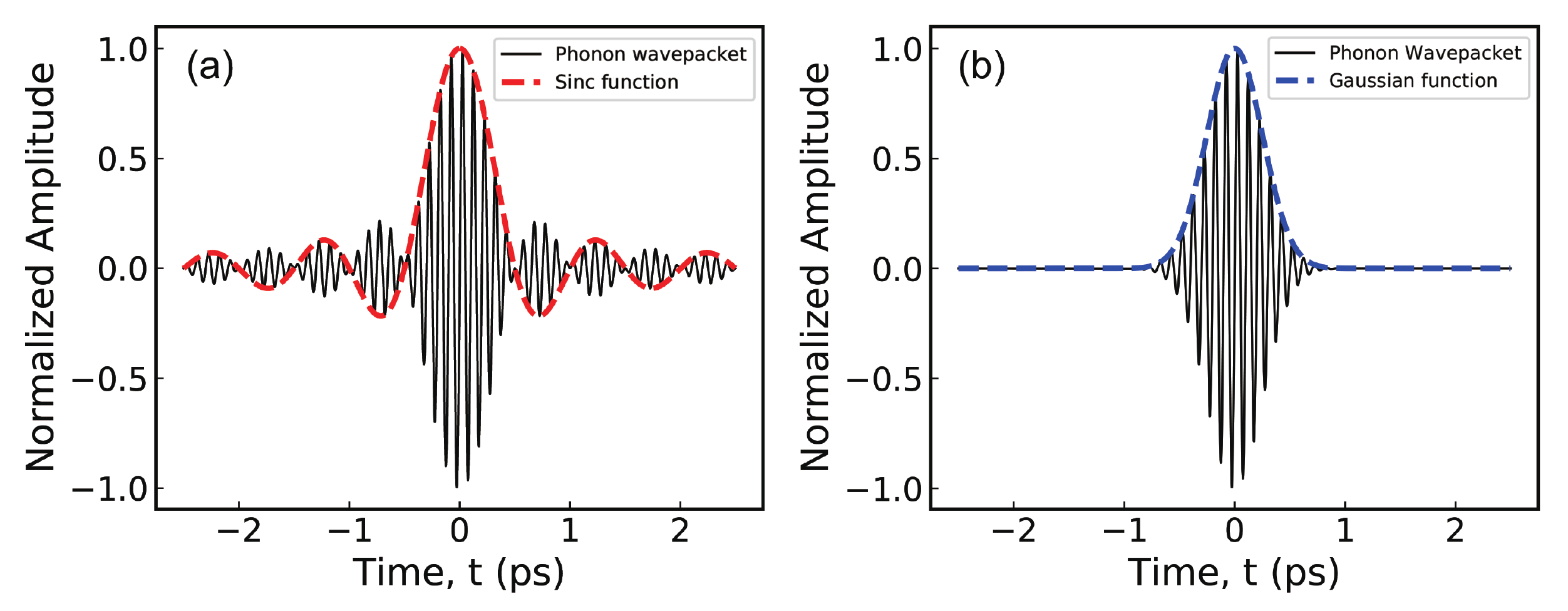}
\caption{ The shape of phonon wavepacket. (a) The shape of the phonon wavepacket in the form of a Sinc function, $e^{-i\omega t} \frac{\sin \pi \Omega t}{\pi \Omega t}$, with $\omega$=10 THz and $\Omega$=1 THz. (b) The shape of the phonon wavepacket in the form of a Gaussian function, $e^{-i\omega t} e^{-A  \Omega^{2}t^{2}}$, with $\omega$=10 THz and $\Omega$=1 THz. The parameter $A$ is set as to 1.96 to make the Sinc and Gaussian functions have equal FWHM.
}
\label{figS1}
\end{figure}

 On the other hand, because the interference occurs when planewaves frequencies are close to each other, $\xi_{\mathbf{k}s{s}'}$ and $\gamma _{\mathbf{k}s{s}'}$ and $g(\omega_{\mathbf{k}'s})$ do not change significantly. We can make a further step for Eq. (\ref{t_16}) as

\begin{eqnarray}
N_{\mathbf{k}s}\left ( t \right )\approx  \bar{\xi}_{\mathbf{k}s}e^{-\bar{\gamma} _{\mathbf{k}s}t}\bar{g}(\omega_{\mathbf{k}s})\int_{-\frac{\Omega_{\mathbf{k}s} }{2}}^{+\frac{\Omega_{\mathbf{k}s} }{2}}e^{-i\Delta\omega_{\mathbf{k}\mathbf{k}'s} t}d\Delta\omega_{\mathbf{k}\mathbf{k}'s},
\label{t_17}
\end{eqnarray}

\noindent where $\bar{\xi}_{\mathbf{k}s}$, $\bar{\gamma} _{\mathbf{k}s}$ and $\bar{g}(\omega_{\mathbf{k}s})$ are respectively the averaged properties over one wavepacket. The above integration yields a sinus cardinal (Sinc) function

\begin{eqnarray}
N_{\mathbf{k}s}\left ( t \right )= 2\bar{\xi}_{\mathbf{k}s}e^{-\bar{\gamma} _{\mathbf{k}s}t}
 \bar{g}(\omega_{\mathbf{k}s})\frac{\sin \pi \Omega_{\mathbf{k}s}  t}{t}.
\label{t_18}
\end{eqnarray}

\noindent where the variables $\bar{\xi}_{\mathbf{k}s}$, $\bar{\gamma} _{\mathbf{k}s}$, $\bar{g}(\omega_{\mathbf{k}s})$ and $\Omega_{\mathbf{k}s}$ are the specific properties of one type of wavepacket. The term $e^{-\bar{\gamma} _{\mathbf{k}s}t}$ corresponds to the time decay of the phonon number, and the term $ \frac{\sin \pi \Omega_{\mathbf{k}s} t}{t}$ indicates the shape of the phonon wavepacket. Compared to a gaussian shape, the sinus cardinal function has longer tails at the both ends, as shown in Fig.\,\ref{figS1}. Note that the time reference $t=0$ corresponds to the time of wavepacket maximum amplitude. This reference will disappear when deriving the autocorrelation of the phonon number. Nevertheless, the two types of wavepackets almost have the same main amplitude at the center. Most importantly, the Eq. (\ref{t_18}) also provides us the information that at time $t$ one wavepacket has emerged and is dynamically varying with the properties $\bar{\xi}_{\mathbf{k}s}$, $\bar{\gamma} _{\mathbf{k}s}$, $\bar{g}(\omega_{\mathbf{k}s})$ and $\Omega_{\mathbf{k}s}$.

Because we consider the gaussian basis in our wavelet transform method and because the gaussian form will also be more tractable in the following, we would like to approximate the sinus cardinal with a gaussian one. The principle is to fit both FWHM which define the temporal coherence. The resulting approximation reads 

\begin{eqnarray}
N_{\mathbf{k}s}\left ( t \right )\approx 4\pi \Omega_{\mathbf{k}s}\bar{\xi}_{\mathbf{k}s}\bar{g}(\omega_{\mathbf{k}s})e^{-\bar{\gamma} _{\mathbf{k}s}t}  e^{-4ln2\cdot \Omega_{\mathbf{k}s}^{2}t^{2}}.
\label{t_19}
\end{eqnarray}

That is, at momemt $t$ the phonon wavepacket with frequency $\omega_{\mathbf{k}s}$ possesses inherent $\bar{\gamma} _{\mathbf{k}s}$ ($\sim $ lifetime) and $\Omega_{\mathbf{k}s}$ ($\sim $ 1/coherence time). Eq. (\ref{t_19}) models the apparition of a single wavepacket. As illustrated in Fig. 1(c), a single wavepacket indeed appears at a given time. To generalize Eq. (\ref{t_19}) to include the description of all wavepackets, the phonon number shall be rewritten as a sum over all wavepacket phonon numbers having each their own peak time $t_{0}$. However, this expression reduces back to Eq. (\ref{t_19}) at a given time. 

\section{\label{sec:level1}Phonon decay}

As was discussed in the main text, the phonon dynamics can be investigated on the basis of the time-autocorrelation of the phonon number. Starting from the Eq. (\ref{t_19}), the autocorrelation function of the phonon wavepacket defined by the physical quantities $\left (\bar{\gamma} _{\mathbf{k}s},\Omega_{\mathbf{k}s} \right ) $ is

\begin{eqnarray}
\left \langle N_{\mathbf{k}s}\left ( t\right )N_{\mathbf{k}s}\left ( 0\right ) \right \rangle  = (4\pi \Omega_{\mathbf{k}s}\bar{\xi}_{\mathbf{k}s}\bar{g}(\omega_{\mathbf{k}s}))^{2} \left \langle \rho_{\mathbf{k}s}\left ( t\right ) \rho_{\mathbf{k}s}\left ( 0\right ) \right \rangle,
\label{p_1}
\end{eqnarray}

\noindent where, $\rho_{\mathbf{k}s}\left ( t\right )=e^{-\bar{\gamma} _{\mathbf{k}s}t-4ln2\cdot \Omega_{\mathbf{k}s}^{2} t^{2}}$. After several derivation steps, the normalized autocorrelation function can be obtained as

\begin{eqnarray}
\frac{\left \langle N_{\mathbf{k}s}\left ( t\right )N_{\mathbf{k}s}\left ( 0\right ) \right \rangle }{\left \langle N_{\mathbf{k}s}\left ( 0\right )N_{\mathbf{k}s}\left ( 0\right ) \right \rangle } = \frac{\left \langle \rho_{\mathbf{k}s}\left ( t\right )\rho_{\mathbf{k}s}\left ( 0\right ) \right \rangle}{\left \langle \rho_{\mathbf{k}s}\left ( 0\right )\rho_{\mathbf{k}s}\left ( 0\right ) \right \rangle}=e^{-\frac{b}{2}t^{2}}\left ( 1-\sqrt{1-e^{-B^{2}}} \right ),
\label{p_2}
\end{eqnarray}

\noindent in which $B=\sqrt{\frac{b}{2}}\left ( t+\frac{a}{b} \right )$, $a=\bar{\gamma} _{\mathbf{k}s}$ and $b=4ln2\cdot\Omega_{\mathbf{k}s}^{2}$. We use the following series expansion to simplify the above equation

\begin{eqnarray}
1-\sqrt{1-e^{-B^{2}}} \approx \frac{1}{2}e^{-B^{2}}-\frac{1}{8}e^{-2B^{2}}+\frac{1}{16}e^{-3B^{2}}-\frac{5}{128}e^{-4B^{2}}.
\label{p_3}
\end{eqnarray}

We now limit the expansion to the first order with $1-\sqrt{1-e^{-B^{2}}} \approx \frac{1}{2}e^{-B^{2}}$ by considering that $B > 1$ in which the first order expansion is accurate. Thus, the normalized autocorrelation function is simplified as

\begin{eqnarray}
\frac{\left \langle N_{\mathbf{k}s}\left ( t\right )N_{\mathbf{k}s}\left ( 0\right ) \right \rangle }{\left \langle N_{\mathbf{k}s}\left ( 0\right )N_{\mathbf{k}s}\left ( 0\right ) \right \rangle } \approx e^{-\frac{b}{2}t^{2}}e^{-B^{2}}= e^{-\bar{\gamma} _{\mathbf{k}s}t}e^{-4ln2\cdot\Omega_{\mathbf{k}s}^{2}t^{2}}.
\label{p_4}
\end{eqnarray}

Previously, it has been demonstrated that the linewidths $\bar{\gamma} _{\mathbf{k}s}$ for mode ${\mathbf{k}s}$ is related to the lifetime $\tau _{\mathbf{k}s}^{l}$, with the relationship

\begin{eqnarray}
 \tau _{\mathbf{k}s}^{l} =\frac{1}{2\bar{\gamma} _{\mathbf{k}s}}.  
 \label{p_5}
\end{eqnarray}

In this work, we define $\tau _{\mathbf{k}s}^{c}$ as the FWHM of the wavepacket. Therefore, by redefining the term $e^{-4ln2\cdot\Omega_{\mathbf{k}s}^{2}t^{2}}$, Eq. (\ref{p_4}) can be rewritten as

\begin{eqnarray}
\frac{\left \langle N_{\mathbf{k}s}\left ( t\right )N_{\mathbf{k}s}\left ( 0\right ) \right \rangle }{\left \langle N_{\mathbf{k}s}\left ( 0\right )N_{\mathbf{k}s}\left ( 0\right ) \right \rangle } = e^{-\frac{t}{2 \tau _{\mathbf{k}s}^{l} }}e^{-4ln2\frac{t^{2}}{{\tau _{\mathbf{k}s}^{c}}^2}}.
\label{p_6}
\end{eqnarray}

In the above equation the exponential part, $e^{-\frac{t}{2 \tau _{\mathbf{k}s}^{l} }}$, results from phonon-phonon scattering. On the contrary, the gaussian part, $e^{-4ln2\frac{t^{2}}{{\tau _{\mathbf{k}s}^{c}}^2}}$, corresponds to the effect of planewave interferences and the formation of wavepackets with a finite duration. Considering the unfolding of phonon number, the autocorrelation function for mode ${\mathbf{k}s}$ along coherence time $\tau _{\mathbf{k}s}^{c}$, $C\left ( t,\tau _{\mathbf{k}s}^{c} \right )$, can be inferred as

\begin{eqnarray}
C\left ( t,\tau _{\mathbf{k}s}^{c} \right )= e^{-\frac{t}{2\tau _{\mathbf{k}s}^{l} }}e^{-4ln2\frac{t^{2}}{{\tau _{\mathbf{k}s}^{c}}^{2}}}. 
\label{p_7}
\end{eqnarray}

\section{\label{sec:level1}Lifetimes and Thermal Conductivities}

\subsection{\label{sec:level2}Lifetimes}

The lifetimes of different modes can be obtained by fitting the autocorrelation function of the phonon energy or of mode velocities. With the MD modal velocities, the normalized autocorrelation function for mode $\mathbf{k}s$ is obtained as

\begin{eqnarray}
C'\left ( t \right )= \frac{ \left \langle \upsilon _{\mathbf{k}s}(t)\upsilon _{\mathbf{k}s}^{\ast}(0) \right \rangle}{\left \langle \upsilon _{\mathbf{k}s}(0)\upsilon _{\mathbf{k}s}^{\ast} (0) \right \rangle},
\label{l_1}
\end{eqnarray}

\noindent $\upsilon _{\mathbf{k}s}(t)$ refers to the time dependent mode velocity for mode $\mathbf{k}s$. Previous references showed that this autocorrelation function can refers to the exponential decay with lifetime ${\tau _{\mathbf{k}s}^{l}}'$,

\begin{eqnarray}
C'\left ( t\right )=e^{-i\omega_{\mathbf{k}s}t}e^{-\frac{t}{{\tau _{\mathbf{k}s}^{l}}'}}.
\label{l_2}
\end{eqnarray}

In this equation, the phonon decay is only considering the phonon-phonon scattering. However, in this work, we find that the phonon decay, i.e. the autocorrelation function, is further coherence. Here, we would like to correct the autocorrelation function by averaging the coherence time dependent autocorrelation functions. By applying the same exponential fitting, we can obtain the lifetime of the corrected autocorrelation function defined as

\begin{eqnarray}
\sum_{\tau _{\mathbf{k}s}^{c}}D\left (\tau _{\mathbf{k}s}^{c} \right )C\left ( t,\tau _{\mathbf{k}s}^{c} \right )=e^{-\frac{t}{{\tau _{\mathbf{k}s}^{l}}}}.
\label{l_3}
\end{eqnarray}

\begin{figure}
[h]
\renewcommand\thefigure{S\arabic{figure}}
\includegraphics[width=0.90\linewidth]{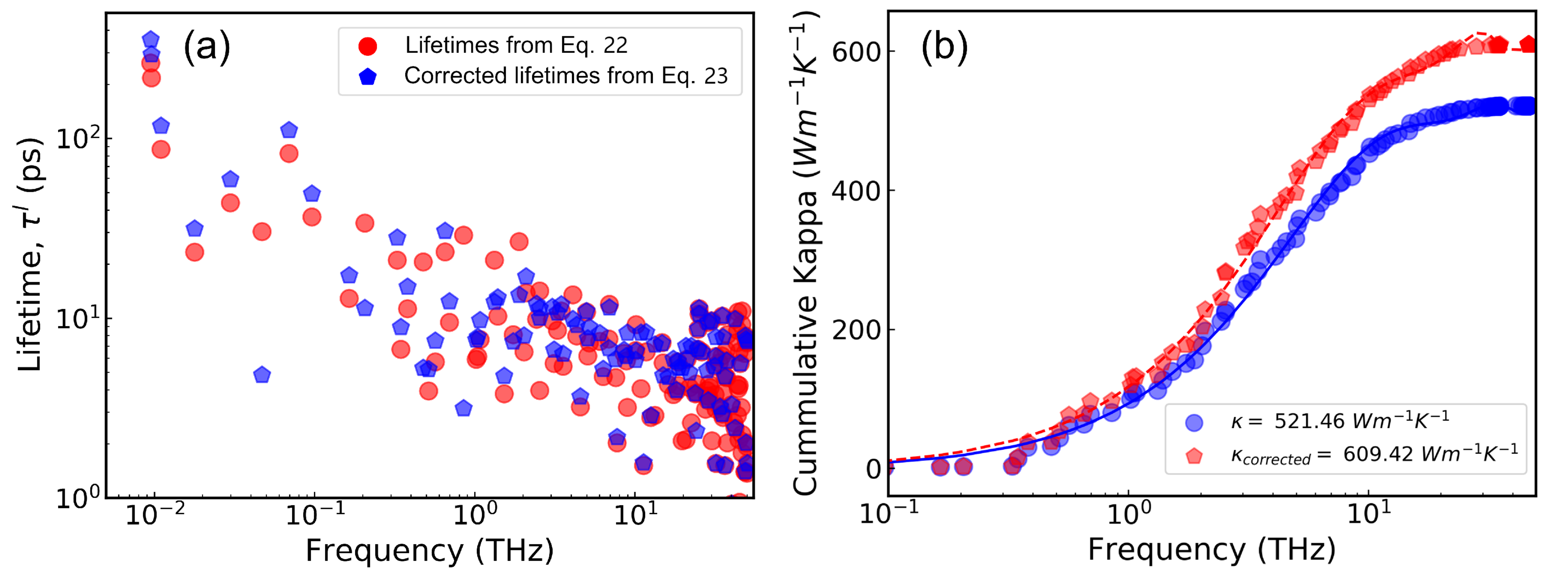}
\caption{(Color online)
The room temperature lifetimes and thermal conductivities in graphene. (a) Room temperature lifetimes of graphene from the fitting of usual autocorrelation function (Eq. (\ref{l_2})) and corrected autocorrelation function (Eq. (\ref{l_3})), respectively. (b) Accumulative room temperature thermal conductivities for graphene based on Eq. (\ref{l_4}).
}
\label{figS3}
\end{figure}

\noindent By fitting the corrected phonon decay function, the obtained ${\tau _{\mathbf{k}s}^{l}}$ time is also corrected by including coherence effects. The comparison between the corrected and the non-corrected lifetimes is shown in Fig.\,\ref{figS3}(a). By including coherence, phonon decay to equilibrium becomes slower, especially at low frequencies. 

\subsection{\label{sec:level2}Thermal Conductivity}

Using the single-mode relaxation time approximation to solve the Boltzmann transport equation, the thermal conductivity can be expressed as

\begin{eqnarray}
\kappa =\sum_{{\mathbf{k}s}}\kappa_{{\mathbf{k}s}}=\sum_{{\mathbf{k}s}}C_{\mathrm{v},{\mathbf{k}s}}\upsilon _{{\mathbf{k}s}}^{2}\tau _{\mathbf{k}s}^{l},
\label{l_4}
\end{eqnarray}

\noindent where the constant volume heat capacity $C_{\mathrm{v},{\mathbf{k}s}}$ for mode $\mathbf{k}s$ can be calculated from $k_{b}\left (  \frac{\hbar\omega }{k_{b}T}\right )^{2}\frac{e^{\frac{\hbar\omega }{k_{b}T}}}{\left ( e^{\frac{\hbar\omega }{k_{b}T}} -1\right )^{2}}$, and $\upsilon _{{\mathbf{k}s}}$ is the group velocity. These phonon infromation are obtained from phonon dispersion based on Lattice Dynamic calculations. Thermal conductivities are calculated with corrected and non-corrected lifetimes. As shown in Fig.\,\ref{figS3}(b), the room temperature thermal conductivity with the non-corrected lifetimes is 521.46 $Wm^{-1}K^{-1}$. This value is lower than the experimental measurement due to the absence of terms from collective interactions, but close to the reported values calculated in the same conditions [4]. Moreover, when including the coherence term, the corrected thermal conductivity reaches 609.42 $Wm^{-1}K^{-1}$, indicating that the wave nature plays a significant role in the phonon transport even at room temperature.

\newpage

\section{\label{sec:level1}Observations in Silicon}

\begin{figure}
[h]
\renewcommand\thefigure{S\arabic{figure}}
\includegraphics[width=0.50\linewidth]{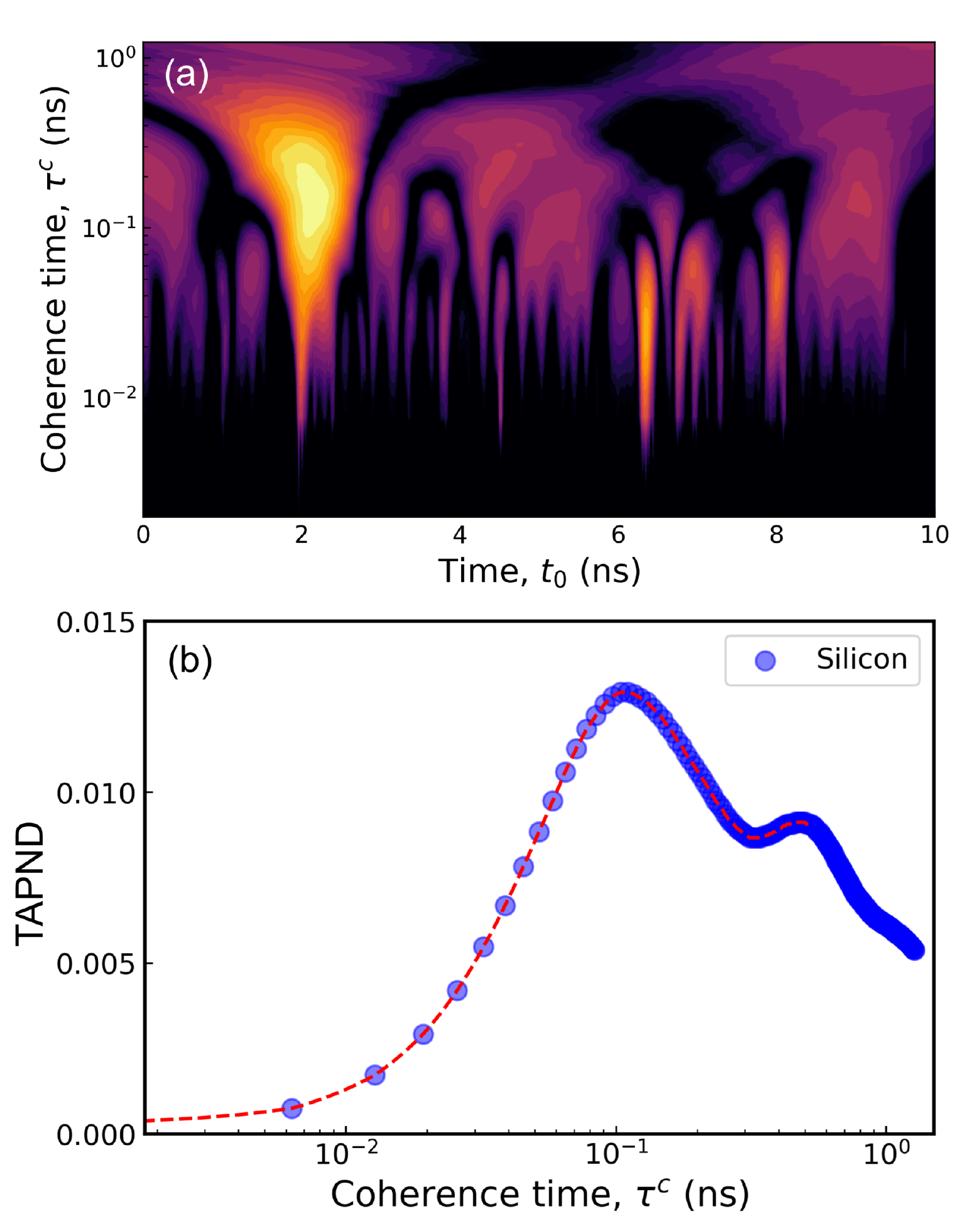}
\caption{Wavelet transform results in silicon at room temperature. (a) Evolution time ($t_{0}$) and coherence time ($\tau _{c}$) dependent phonon number density for the $0.05M$ TA mode of Silicon at room temperature. (b) Time-averaged phonon number density (TAPND) as a function of coherence time for the $0.05M$ TA mode at room temperature.
}
\label{figS4}
\end{figure}

[1] J. M. Larkin, J. E. Turney, A. D. Massicotte, C. H. Amon, and A. J. H. McGaughey, J. Comput. Theor. Nanosci. $\mathbf{11}$, 249
(2014).

[2] R. J. Hardy, Phys. Rev. $\mathbf{132}$, 168 (1963).

[3] A. J. C. Ladd, B. Moran, and W. G. Hoover, Phys. Rev. B $\mathbf{34}$, 5058 (1986).

[4] S. Srinivasan and G. Balasubramanian, Langmuir $\mathbf{34}$, 3326 (2018).